# NEUTRINO-FLUX VARIABILITY, NUCLEAR-DECAY VARIABILITY, AND THEIR APPARENT RELATIONSHIP

P.A. Sturrock[1]

[1] Center for Space Science and Astrophysics and Kavli Institute for Particle Astrophysics and Cosmology, Stanford University, Stanford, CA 94305-4060, USA; sturrock@stanford.edu

**ABSTRACT**

Analysis of Homestake, Gallex and GNO measurements reveals evidence of variability of presumed solar-neutrino-flux measurements. Analysis of Super-Kamiokande neutrino records over the interval May 1996 to July 2001 reveals oscillations at 9.43 year$^{-1}$ and 12.6 year$^{-1}$, both well within a range of frequencies (6 – 16 year$^{-1}$) that, according to helioseismology, could be related to internal solar rotation.

Analysis of the results of a nuclear-decay experiment carried out at the Brookhaven National Laboratory over the time interval 1982 – 1986 reveals a strong annual oscillation and also strong oscillations at 11.2 and 13.2 year$^{-1}$, both of which would, according to helioseismology, be compatible with influences of internal solar rotation. Similar oscillations are found in an extensive series of nuclear-decay measurements conducted by Alexander Parkhomov of the Lomonosov Moscow State University and the Russian Academy of Natural Sciences. By contrast, as noted by Stefan Pomme of the European Commission Joint Research Centre and his colleagues, nuclear-decay measurements acquired at standards laboratories tend not to exhibit evidence of variability.

The most extensive series of nuclear-decay measurements comes from an experiment initiated by the late Gideon Steinitz at the Geological Survey of Israel. This experiment, which was in operation from January 2007 to November 2016, recorded 340,000 lines of radon-related measurements from three gamma detectors and three environmental detectors (temperature, pressure, and line voltage). Analysis of a subset of 85,000 lines of hourly gamma measurements reveals overwhelmingly strong evidence of diurnal, annual and semi-annual oscillations and a number of oscillations with frequencies compatible with influences of internal solar rotation. There is no correlation between the gamma measurements and the environmental measurements.

The rotational modulations may be attributed to an influence of the solar internal magnetic field by the RSFP (Resonant Spin-Flavor Precession) process. The detection of several pairs of oscillations separated by precisely 1 year$^{-1}$ may be attributed to misalignments of internal rotation axes with respect to the normal to the ecliptic. A triplet of oscillations (at effectively 7.43, 8.43 and 9.43 year$^{-1}$) may be attributed to an internal region (presumably the core) that has a sidereal rotation rate of 8.43 year$^{-1}$ and a rotation axis approximately *orthogonal* to that of the solar photosphere. These results suggest that the Sun had its origin in more than one stage of condensation of interplanetary material (one on top of another), which would presumably lead to layers of the solar interior that have different metallicities, as well as different rotation rates and axes.

It is remarkable that the oscillation at 9.43 year$^{-1}$ occurs in both Superkamiokande and GSI data with the same amplitude and the same phase.

Analysis of GSI data, together with a review of experiments conducted by Enrico Bellotti and his collaborators of the Instituto Nazionali di Fisica Nucleare, suggests that neutrinos do not influence decay rates, but do influence – presumably by a collective process - the direction of emission of decay products. This can help explain why the GSI experiment – for which decay products travel through air – gives evidence of strong modulation, whereas experiments at standards laboratories – for which decay products typically travel through comparatively dense media – do not.

The peak modulation occurs near local midnight in early June, suggestive of a role of cosmic neutrinos. These neutrinos could provide the mass attributed to dark matter for a neutrino mass of order 0.1 eV.

# 1. Introduction

This review examines the following questions:
1. What do we know concerning the variability of neutrino fluxes, focusing primarily on the solar neutrino flux but considering also a possible cosmic flux?
2. Is the beta-decay process constant or is it variable?
3. If the beta-decay-related flux is variable, is there evidence for an influence of neutrinos?

We inquire into the variability of the solar neutrino flux as measured by the Homestake experiment in Section 2; as measured by the GALLEX, GNO and SAGE experiments in Section 3; and as measured by the Super-Kamiokande Neutrino Observatory in Section 4.

In Section 5, we review some of the more detailed early evidence for beta-decay variability, beginning with an experiment conducted at the Brookhaven National laboratory (BNL) over the time interval 1982 to 1986 (Alburger et al., 1986).

In Section 6, we review the most detailed evidence now available concerning beta decay variability. This evidence comes from radon-decay measurements made at the Geological Survey of Israel (GSI; Steinitz et al., 2011, 2014).

In Section 7, we review two important experiments by Enrico Bellotti. We discuss the apparent role of cosmic neutrinos – and their possible relationship to dark matter – in Section 8 and discuss our overall finding in Section 9.

Appendix A presents a likelihood procedure (related to the familiar Lomb-Scargle procedure (Lomb (1976), Scargle (1982)) for computing the power, amplitude and phase of irregularly spaced measurements, discrete or binned.

Appendix B briefly reviews relevant publications from standards laboratories.

Appendix C lists some proposed experiments

## 2. Homestake

The Italian nuclear physicist Bruno Pontecorvo pointed out in 1946 (when he was working at the Chalk River laboratory in Canada) that a neutrino of energy 0.814 MeV or more may be captured by an atom of $^{37}$Cl, converting it into $^{37}$Ar which will decay back to a chlorine atom in (on average) 35 days, emitting a 2.8 keV electron. In 1964, the chemist Raymond Davis at the Brookhaven National Laboratory and the physicist John Bahcall at the California Institute of Technology proposed an experiment to detect solar neutrinos by that mechanism (Bahcall, 1964; Davis 1964).

The technical challenges were enormous. To start with, it was necessary to locate the experiment deep underground to avoid cosmic rays. Davis was able to negotiate the use of space 4,850 feet underground in the Homestake gold mine in Lead, South Dakota. This meant that every piece of equipment had to be small enough to fit into the cage of the mineshaft, which was no mean challenge, since the main part of the detector was to be a chamber containing 615 tons of perchlorethylene, the fluid typically used for dry-cleaning. (Bahcall used to joke that if the neutrino experiment did not work, they could always go into the dry-cleaning business.) The experiment was in operation from April 1970 until May 1994.

Bahcall and others had made extensive calculations of the number of neutrinos that one could expect the experiment to capture in course of a "run," which would typically last three or four weeks. These challenging calculations were dependent on assumptions concerning the internal composition and structure of the Sun, and on laboratory measurements relevant to the processes involved in the production and detection of neutrinos. The detector would be sensitive to only a small fraction of the neutrinos produced in the core since most solar neutrinos have energies close to 233 keV, whereas the chlorine experiment is sensitive only to neutrinos with energy close to 814 keV.

When Davis's Homestake experiment was finally up (and down) and running, there was a problem. The prediction had been that the experiment should detect about 8 neutrinos per day. The experiment actually measured only about one neutrino every second day. To measure this low a flux meant that it was necessary to extract from the 600 tons of fluid only a few atoms of radioactive argon—and then count them! New measurements of nuclear cross-sections and new calculations brought the theoretical estimate down to a little over one neutrino per day, but there remained an irreducible discrepancy—Homestake measured only about one-third as many neutrinos as expected.

It took a few years for the discrepancy to be accepted as real, but in 1976 Bahcall and Davis published an article entitled *Solar Neutrinos: A Scientific Puzzle* (Bahcall and Davis, 1976). This article sets out the contradiction between the theoretical expectation and the experimental results. A second neutrino experiment began operation in July 1983 and was found to give similar results. This was the Japanese Kamiokande experiment, so named because it is located in a mine of the Kamioka Mining and Smelting Company at Kamioka in Hida Province, 300 km west of Tokyo in the Japanese Alps. The Kamiokande experiment, which detected Cerenkov radiation from electrons that had been hit (and accelerated) by neutrinos, detected about one half as many neutrinos as were expected. These two anomalous results led physicists to begin a period of intense study to determine whether the discrepancies were due to mistaken solar physics or mistaken particle physics.

This possibility had led several physicists to compare the measured neutrino flux with various measures of solar activity. For instance, Bieber et al. (1990) and Bahcall and Press (1991) compared the Homestake measurements with the sunspot number. Both analyses gave evidence suggestive of an association, but the authors were suitably cautious. When measurements by Kamiokande and later Super-Kamiokande failed to show evidence for a solar-cycle variation, the physics community concluded that the early analyses of Homestake data had—for whatever reason—been misleading, and that there is no real connection between the neutrino flux and the solar cycle.

However, that may have been too early a judgment. With the cooperation of Guenther Walther, Professor of Statistics at Stanford University, Sturrock and Wheatland carried out a power-spectrum analysis of the Homestake data, leading to the power spectrum shown in Figure 1. This shows peaks at 10.83, 14.85, 12.88, 13.85, and 14.88 $y^{-1}$. This group of frequencies is suggestive of modulation of the solar neutrino flux in the radiative zone, rather than the convection zone (which is the seat of the sunspot cycle). Modulation of the neutrino flux could occur in the radiative zone by the Resonant Spin Flavor Precession (RSFP) mechanism (Akhmedov, 1988; Pulido, Das & Picarello 2010), which involves magnetic field, changing the flavor of a neutrino from one form to another. The array of oscillations, separated by 1 $y^{-1}$, could be attributed to a dependence of the modulation on solar latitude. However, later research (concerning r-mode oscillations), points to the alternative possibility that the 1 $y^{-1}$ offset in the frequencies may be attributable to an "oblique rotator" effect, which could occur if the axis of rotation of the radiative zone differs from the normal to the ecliptic (Sturrock and Bai, 1992). We shall examine this possibility further in Section 5.

## 3. GALLEX and GNO

These experiments were designed to detect $^{71}$Ge nuclei resulting from the transition $^{71}$Ga(p,n)$^{71}$Ge (which has the low threshold of 233 keV) in liquid gallium metal. The GALLEX Experiment was in operation from 1991 to 1997 (Anselmann et al. 1993, Anselmann et al. 1995; Hampel et al 1996); Hampel et al 1999), and the closely related follow-on GNO experiment was in operation from 1998 to 2000 (Altman et al. 2000). These experiments were sensitive to the more abundant 233 keV neutrinos, compared with the 814 keV neutrinos detected by Homestake.

Since the GALLEX and GNO experiments were very similar, these two datasets have been analyzed together by Sturrock and Weber (2002) and by Pandola (2004). The flux versus time, from both analyses, is shown in Figure 2.

These articles each show a power spectrum over the frequency range 10 - 20 y$^{-1}$, derived by the Lomb-Scargle procedure (Lomb, 1976, Scargle 1982). The results of the analyses are very similar. The Sturrock-Weber version of the power spectrum, derived from an analysis of measurements assigned to the mean time of each run, is shown in Figure 3. The strongest oscillation in that band is found at 13.6 y$^{-1}$, with power S = 5.93. It is interesting that this corresponds to the sidereal rotation frequency of the radiative zone as determined by helioseismology (Schou et al., 1998).

The probability of finding a power S by chance at a specified frequency, if the data are in fact derived from a normal (Gaussian) distribution, is given (Scargle, 1982) by

$$P = e^{-S}.$$
(1)

If the oscillation is not specified but is known to be one of M possibilities, the probability of this result - known (Scargle, 1982) as the "false alarm" probability - is given by

$$FAP = 1 - \left(1 - e^{-S}\right)^M .$$

(2)

We see from Equation (1) that the probability of finding by chance a power $S = 5.0$ at that particular frequency, is less than 1%.

However, finding an oscillation at the sidereal rotation frequency of the radiative zone is somewhat surprising. One would expect to find an oscillation that originates in the radiative zone at the corresponding synodic frequency (12.6 $y^{-1}$), the frequency as it would be determined from observations on Earth. However, this expectation is based on the assumption that the rotation axis of the radiative zone is close to the normal to the ecliptic (similar to that of the convection zone, as determined by observation of the photosphere). Hence this discrepancy calls into question the assumption that there is a single rotation axis of the entire solar interior that is close to the normal to the ecliptic and therefore close to the rotation axis of the planetary system. We shall have further comments on this issue in Section 6.

## 4. Super-Kamiokande

The Super-Kamiokande Consortium has provided an extensive compilation of solar neutrino measurements, on a timescale of minutes (with one break due to an accident). Measurements acquired from May 31, 1996, to July 15, 2001, organized into 184 10-day bins and 358 5-day bins, were published by the Super-Kamiokande Consortium in 2003 (Yoo et al., 2003). The abstract reads in part: *We employed the Lomb test to look for periodic modulations of the observed solar neutrino flux. The obtained periodogram is consistent with statistical fluctuation and no significant periodicity was found.*

According to figures in that article, the Consortium scanned a very wide frequency range (up to 70 year$^{-1}$). The maximum power appears to have been 7.8. The probability of finding a power $S$ or more at a specified frequency is $e^{-S}$ (Equation (1)), i.e. 0.0004. However, for a dataset of duration $T$, the Nyquist frequency is $1/2T$, which for the Super-Kamiokande dataset is 0.1 year$^{-1}$. Hence there are 700 independent frequency bins in the search band of width 70 year$^{-1}$ considered by Yoo et al., so that the probability of finding a peak of power 7.8 or more anywhere in that band is 0.3, which is not significant. On the other hand, if one were to limit one's search to evidence of solar rotation, a search band of width 10 year$^{-1}$ (e.g. from 6 year$^{-1}$ to 16 year$^{-1}$), as may be derived for instance from Schou et al. (1998), would be reasonable, so that the probability of finding a peak of power 7.8 anywhere in that band would be 2%, which is not altogether insignificant.

More to the point, *Yoo et al. did not take account of all the relevant available information*. For each bin, the tabular record provides the start time, end time, the lower error estimate and upper error estimate, as well as the measured neutrino flux. Koshio (2003), also a member of the Super-Kamiokande Consortium, carried out a more detailed analysis of the Super-Kamiokande measurements, taking account of the width of each time bin, but still using a very wide search band (50 year$^{-1}$). In this analysis, the principal peak in the power spectrum is again found at 9.43 year$^{-1}$, now with a power of 10.88. Noting that there are 500 independent peaks in a search band of 50 year$^{-1}$, this power is significant at the 1% level. For a search band of 10 year$^{-1}$, which would be more appropriate for a study of rotational modulation, this peak would be significant at the 0.2% level.

The Super-Kamiokande measurements have also been analyzed by Milsztajn (2003), and also by Nakahata (2003) for the Super-Kamiokande Consortium. The results of these analyses, and of variants of these analyses, are shown in Sturrock et al. (2005). Table 1 lists the power at 9.43 year$^{-1}$ for five analyses, including a more recent analysis (Sturrock and Scargle, 2006) to be discussed later. We see that the power tends to increase as the analysis takes account of more experimental information, which is what one would expect of a real physical oscillation. It is also worth noting that the second-strongest oscillation in the rotational search band has a frequency close to 12.3 year$^{-1}$, the synodic rotational frequency of the solar radiative zone. *This suggests that modulation of the solar neutrino flux occurs deep inside the Sun, not at the surface or in the convection zone.* It may be noted that, on this basis, one would not expect to find any association between the solar neutrino flux and solar activity, which is believed to occur primarily in the outer layers of the Sun, including the corona.

A likelihood analysis that takes account of all of the currently available Super-Kamiokande information (Sturrock & Scargle, 2006), including the asymmetry in the error estimates, leads to a power spectrum, shown in Figure 4. The top 10 peaks in the frequency band 6 – 16 year-1, are listed in Table 2. The maximum power, 13.24, is found at 9.43 year$^{-1}$. The probability of finding a peak of this power or more in a search band of width 10 year$^{-1}$ is 0.02%.

Extensive analysis of Super-Kamiokande neutrino measurements (for both 10-day bins and 5-day bins) has been carried out by Ranucci (2006), using likelihood procedures. For an analysis of the 10-day-bin dataset, the two strongest peaks are found at 9.42 year$^{-1}$ and 26.57 year$^{-1}$. Ranucci points out that one of these oscillations must be an alias of the other, since the acquisition schedule has a periodicity with frequency 36 year$^{-1}$, and 9.42 + 26.57 ≈ 36. The peak at 26.57 year$^{-1}$ is not present in the 5-day data (as one would expect if it were merely an alias).

**11**

For a "weighted periodogram" analysis, Ranucci find a power of 10.84 at 9.42 year$^{-1}$, which is indistinguishable from 9.43 year$^{-1}$. Ranucci also presents an analysis that takes account of the asymmetry in the error estimates, but arrives at a somewhat smaller power (9.42).

A more recent likelihood analysis of the Super-Kamiokande measurements (using the procedure of Appendix A) determines not only the power, but also the amplitude and phase of each oscillation (Sturrock et al., 2020). This extra information will be found useful when we come to compare solar neutrino oscillations with oscillations in nuclear decay measurements.

Ranucci and Sello (2007) have analyzed Super-Kamiokande data by means of the wavelet procedure (Grossman & Mortlet, 1984). This procedure has the convenient property of analyzing data in terms of both time and frequency as a single operation, a procedure that works well when the resulting structure has a small number of discrete components. Figure 5 reproduces their wavelet map of the 5-day binned Super-Kamiokande I data for the period range 0.02 y to 5.1 y. This reveals oscillations with periods 2.8 y and 1.1 y, both significant at the 5% confidence limit, but not at the 1% limit. An oscillation at 9.43 year$^{-1}$ (period 0.106 y) is not evident in this display. This region of the wavelet map is seen to be highly fragmented.

Ranucci and Sello also compared their analysis of SK and SNO data with both solar magnetic-field measurements compiled by the Wilcox Solar Observatory (WSO; http://quake.stanford.edu/wso) and with solar-flare activity data compiled by Osguc and Atac (1989). However, the WSO instrument records the weak "background" photospheric magnetic field, not the strong active-region magnetic field that is responsible for solar activity. It should also be noted that solar-flare activity occurs primarily in the solar corona – not in the photosphere. Ranucci and Sello concluded that *the 9.42 – 9.43 yr$^{-1}$ component is likely to be only a peculiar feature of the SK data, with no connection to the underlying neutrino flux.* However, we shall find in the next section that the oscillation at 9.43 year$^{-1}$ is clearly evident in other datasets, indicating that it represents a true solar feature.

Attempts to relate apparently anomalous neutrino measurements to solar activity (Jenkins et al., 2009) have not been successful, although the topic is still of interest (Mohsinally, 2016).

Calculations by Lingenfelter et al. (1985) have shown that the generation of neutrinos by solar flares is far too weak to be detectable by current neutrino experiments. They estimate that the neutrino production by even a large flare - comparable to those of August 2 and 7, 1972 - would be too small by ten orders of magnitude to be detectable by Homestake or a similar experiment.

We show in Figure 6 a spectrogram formed from Super-Kamiokande data for the frequency band $6 - 16$ year$^{-1}$. Spectrograms show - more clearly than wavelets - how oscillations vary with time. We see that oscillations at 8.4 year$^{-1}$, 9.4 year$^{-1}$, and 12.6 year$^{-1}$ are all steady for the year 1988, but are less obvious after that date. We infer that the process that leads to oscillations in the neutrino flux - presumably the RSFP (Resonant Spin Flavor Precession) operation (Akhmedov, 1989; Akhmedov, Lanza and Petkov, 1995) - is time varying, reflecting variations in the relevant magnetic field, i.e. the magnetic field in the radiative zone.

## 5. Early Evidence for Beta-Decay Variability

There has for some time been evidence that some beta decay processes exhibit some form of variability. Whether or not beta decays are intrinsically variable is significant for geologists who rely on radon measurements to probe the outer layers of the lithosphere. Whether or not the solar neutrino flux is variable is important not only to solar physicists, but also to physicists for whom solar-neutrino measurements yield a test of our understanding of nuclear physics.

David Alburger and his colleagues at the Brookhaven National Laboratory have reported the results of their study of the decay of $^{22}$Si over the time period 1982 to 1986, using the long-lived nuclide $^{36}$Cl as a calibration standard (Alburger et al., 1986). They noted *small periodic annual deviations of the data points from an exponential decay curve* (that were) *of uncertain origin*. It is notable that the depths of modulation – of order 0.05% – for the two nuclides are similar, even though there is a wide difference in the decay half-lives (172 y for $^{32}$Si, 300,000 y for $^{36}$Cl).

The Alburger data have since been further analyzed (Sturrock, Buncher, Fischbach, et al., 2009). The ratio of the normalized $^{32}$Si measurements and the normalized $^{36}$Cl measurements is shown, as a function of time, in Figure 7. The annual oscillation is obvious. A power spectrum analysis of this ratio is shown in Figure 8. As expected, there is a strong peak (with power $S = 24$) at $v = 1$ y$^{-1}$. According to Equation 1, the probability that this peak has occurred by chance at the annual frequency is $4\ 10^{-11}$.

The next most significant peak is found at frequency 11.17 y$^{-1}$ with power 20.76. If this frequency had been chosen in advance, the probability of finding this peak by chance would be $10^{-9}$. However, since the duration T is 7 years, the Nyquist frequency

$$v_{Ny} = \frac{1}{2T}$$

(5-2)

is found to be 0.064 y$^{-1}$. Hence there are 20/0.064, i.e. 156, independent candidate frequencies in the band 6 – 16 y$^{-1}$. This leads to a probability of $1.5\ 10^{-7}$ of finding a peak of power $S = 20$ by chance anywhere in the rotational search 6 - 16 y$^{-1}$. (An oscillation at or near 11.2 y$^{-1}$ will be found in other datasets.)

Hence, not only is the annual oscillation in the BNL data highly significant, but so also is the oscillation at 11.17 y$^{-1}$. Since this frequency is significantly lower than the synodic rotation frequency of the convection zone (which is the source of solar activity), this raises the possibility that this observation may have its origin in internal solar processes, the interior rotating *more slowly* than the visible surface.

As we discuss in Appendix B, the question of whether nuclear decay rates are constant or variable is of great interest to analysts at standards laboratories. Helmut Siegert and his colleagues at the Physikalisch Technische Bundesansalt in Braunschweig, Germany, have reported the results of a 20-year study of the beta decays of $^{152}$Eu and $^{154}$Eu, using $^{226}$Ra as a standard (Siegert et al., 1998). The $^{226}$Ra reference source was a 40-year-old specimen of solid radium sulphate that was sealed in a stainless steel tube. Siegert et al. noted annual variations in the measured decay rates of both $^{152}$Eu and $^{226}$Ra. They suggested that *The oscillation may be explained by a discharge effect on the charge-collecting capacitor, the cables, and the insulator to the ionization chamber electrode, caused by a background radioactivity such as radon and daughter products which are known to show seasonal concentration changes.*

Eckhart Dieter Falkenberg, of Uhldingen, Germany, claimed to find evidence of an annual oscillation in the beta decay of tritium, which he attributed to the annual variation of the Earth-Sun distance, suggesting a possible role of neutrinos (Falkenberg, 2001).

Yuriy Alexeevich Baurov and his collaborators at the Research Institute of Cosmic Physics in Moscow have carried out investigations of the variability of beta-decay rates of radioactive elements for over a decade (see, for instance, Baurov et al. 2001, 2013). They typically found a strong diurnal oscillation that has a shape characteristic of caustics. They also found an association between the strength of the effect and right ascension, suggesting a cosmological origin of the variation. Since he was considering a possible cosmic influence on nuclei, Baurov et al. expressed their conjectured interpretation in terms of a hypothetical *cosmic vector* *potential*. Baurov et al. also identified an oscillation with a period of about 27 days, suggestive of a solar influence.

Alexander Parkhomov, of the Lomonosov Moscow State University in Moscow, has investigated nuclear decays and related phenomena for many years (Parkhomov, 2005, 2010a, 2010b, 2010c, 2010d, 2011), recently summarizing his research in Parkhomov (2019). Parkhomov, who suspects that there may be a cosmic influence on beta decays, has found evidence of variability in beta decays but not in alpha decays.

Ephraim Fischbach of Purdue University and his colleagues published a review article in 2009 (Fischbach et al., 2009). This article presents an overview of research - up to that date - *concerning the question of whether nuclear decay rates are time-independent constants of nature or parameters that can be altered by an external perturbation.* It was at that time reasonable to assume that variations in flux measurements should be interpreted as variations in decay rates. It was also reasonable to consider that an annual variation in these measurements may be related to the annual variation in the Earth-Sun distance. However, we shall find that information acquired since the date of that article leads us now to question those assumptions.

## 6. The Geological Survey of Israel Radon Measurements

The most extensive dataset concerning beta-decay variability is one derived from a radon experiment at the Geological Survey of Israel (GSI), which was in operation from January 28, 2007 to November 7, 2016 (Steinitz et al. 2011, 2014). This dataset tabulates measurements of gamma and alpha radiation from the beta-decay of radon, and also measurements of temperature, pressure, and supply voltage, these measurements all being recorded every 15 minutes. We here examine a compressed version, comprised of approximately 85,000 hourly measurements. As we shall find from Section 7, a crucial (and unanticipated) characteristics of this experiment is that the decay of radon, and the propagation of the decay products, occurred entirely in air.

Since there is a very strong diurnal variation, it has proved advantageous to examine beta-decay measurements as a function of hour of day. Figures 9 and 10 show power spectra formed from measurements acquired at both local noon and local midnight over the frequency ranges 0 - 6 year$^{-1}$ and 6 - 16 year$^{-1}$, respectively. Figure 9 shows that, for noon measurements, the strongest oscillation is that at 1 year$^{-1}$ but Figure 10 shows that, for midnight measurements, the strongest oscillation is one at 2 year$^{-1}$. (This already calls into question the expectation that an annual oscillation is necessarily attributable to the varying Earth-Sun distance.)

The principal oscillations for the frequency range 6 – 16 year$^{-1}$ are listed in Table 3 for the noon measurements and in Table 4 for the midnight measurements.

Analysis of the 21,000 "noon" measurements, acquired between 10 am and 2 pm, reveals a strong annual oscillation with power S = 4250. According to the standard formula for significance estimation, $P = e^{-S}$, there is a probability of only $10^{-1846}$ (virtually impossible) of finding an oscillation with that power or more at a specified frequency for randon measurements modeled as a Gaussian distribution. For midnight data acquired between 10 pm and 2 am, the strongest oscillation is the semi-annual oscillation for which the power is 2020, for which the standard formula leads to a probability estimate of $10^{-877}$ (again, virtually impossible) of finding an oscillation with that power or more at a specified frequency for random measurements modeled as a Gaussian distribution.

However, the midnight measurements (but not the noon measurements) also show a sequence of oscillations in a frequency band appropriate for an influence of internal solar rotation. The implication of this finding is that radon measurements are influenced by some form of solar radiation which would be traveling vertically upwards at midnight. Since that radiation would have traveled (virtually unimpeded) through the Earth, this fact strongly supports Falkenberg's conjecture that nuclear decays are somehow influenced by solar neutrinos.

These oscillations in the midnight measurements indicate that the neutrino flux is modulated by some process that occurs in the solar interior. A prime possibility for this process is the RSFP (Resonant Spin Flavor Precession) process whereby a neutrino traveling through a magnetized plasma may change its flavor (Akhmedov 1988; Pulido et al., 2010).

For a dataset of length 10 years, we expect to be able to detect 200 independent oscillations in a band of width 10 year$^{-1}$. Table 4 lists the frequencies and powers of the top 20 peaks in the power-spectrum formed from midnight measurements over the frequency band 6 - 16 year$^{-1}$, which covers the frequency band of internal solar rotation as determined from helioseismology (Schou et al., 1998).

The oscillation at 12.63 year$^{-1}$ is of special interest since that corresponds to the *synodic* rotation frequency (the frequency as observed from Earth) of the solar radiative zone. The oscillation at 13.67 year$^{-1}$ corresponds to the *sidereal* rotation frequency (the frequency as observed from space). The presence of this pair of oscillations suggests that the axis of rotation of the radiative zone departs significantly from the normal to the ecliptic (Sturrock & Bai, 1992).

It is notable that this power spectrum contains a peak at 9.43 year$^{-1}$, exactly the frequency that is most evident in the power spectrum formed from Super-Kamiokande data. It is also notable that this oscillation has been detected by a quite different experiment at the Baksan Neutrino Observatory (Alexeyev et al., 2018).

Although we have here concentrated on articles by Steinitz and his collaborators that are relevant to solar physics, they have published many other articles concerning radon beta decays. Two articles (Steinitz et al. 2015, 2018) exhibit clear evidence that oscillations detected in radon beta-decay data are not specific to any particular environment. The former describes sub-surface measurements made by an experiment in Tenerife in the Canary Islands. The latter describes measurements of both radon and thoron made in Arad, Romania, which exhibit patterns found also in GSI measurements.

Two other GSI articles (Steinitz et al. 2013, 2015) are of special interest in showing that radon beta-decay measurements exhibit *directional* characteristics. The latter describes an experiment with a goniometer-like design, involving two distinct "channels" for detecting gamma rays that are at right angles to each other. The two sets of measurements were completely different, showing that *directionality plays a key role in beta-decay studies*. Another experiment (Steinitz et al., 2016) is unique in exhibiting measurements that are clearly related to man-made (currently unidentified) operations.

## 7. Two Bellotti experiments and their implications

We now review two highly significant experiments by Bellotti et al. (2015).

The first experiment examined decay products from radon contained in air inside a spherical container. We show in Figure 11 a 5-day section of the measurements, exhibiting an unmistakable diurnal variation.

For comparison, we show in Figure 12 a similar normalized 5-day sample of GSI gamma measurements. We see that the variation is very similar to the variation of the first Bellotti experiment shown in Figure 11. For both plots, the minimum to maximum excursion is in the range 7 - 8 %, and both plots show both a major and a minor peak in each daily section. There is so close a similarity that one must surely suspect that they are sampling the same stimulus. Since we saw in Section 6 that GSI gamma measurements exhibit variations that could be attributed to an influence of solar neutrinos, the similarity of the plots shown in Figures 11 and 12 suggests that the diurnal variations apparent in both the Bellotti experiment and the GSI experiment are also attributable to an influence of solar neutrinos.

In the second experiment, radon in air was replaced by radon in olive oil. *In this experiment, there was no evidence of a diurnal variation.* The introduction of olive oil obviously has a major influence on the experiment, but it can hardly be influencing either the solar neutrino flux or the nuclear decay process. Its role can be understood as that of converting a collimated distribution of photons into an isotropic distribution. This suggests that *the influence of neutrinos is not to determine whether or not decay occurs, but to influence the photon distribution if and when decay occurs.*

This proposition is supported by a dedicated experiment by Steinitz, Kotlarsky & Piatibratova (2015), which showed explicitly that measurements of decay products by two detectors at the same location, each of which was responsive to only a discrete range of solid angle, varied with the directional sensitivity of the two detectors.

These results suggest that, for any experiment dedicated to beta-decay measurements which exhibit variability in the measurements, there may actually be no variation in the nuclear decay rate (as it would for instance be characterized by a half-life). *Any variability may be attributable instead to variation in the direction of propagation of decay products*.

We may infer that *the directional characteristics of emission are preserved if propagation occurs purely in air, but the directional characteristics will be modified and perhaps lost if propagation occurs in a non-transparent medium.*

This interpretation can explain why many experiments carried out at standards laboratories fail to show evidence of variability. In these experiments, decay products typically travel - in whole or in part - through a photon-scattering medium, namely the "cocktail" in which a specimen is encased and the porcelain vial in which the "cocktail" is contained. Converting a photon distribution that has a well-defined directionality into one that is isotropic effectively removes the signal that was originally coded in the directionality.

An experiment by Steinitz et al. (2015) independently establishes that whatever may be responsible for influencing a beta-decay process is intrinsically directional. They found that two detectors, at the same location, each of which was responsive to only a discrete range of solid angle, gave different measurements if the detectors were sensitive to photons traveling in different directions.

The role of directionality will be discussed further in the Discussion section (Section 9).

## 8. Cosmic Neutrinos

Alexander Parkhomov, of the Russian Academy of Natural Sciences and the Lomonosov Moscow State University, has carried out research concerning nuclear decays for many years, finding evidence of variability of beta decays but not of alpha decays (Parkhomov, 2010a, 2011).

Parkhomov (2018) has drawn attention to the possibility that beta decays may be influenced by "cosmic slow neutrinos," as well as by solar neutrinos. Parkhomov also describes experiments in which some form of radiation, that has an influence on beta decays, can be detected by a "telescope" that has a parabolic steel mirror. Parkhomov raises the possibility that such slow neutrinos may be a contributor to dark matter. We explore this possibility in this section.

Figure 13 shows the normalized beta-decay signal as a function of date and of hour of day. According to this figure, the dominant contribution to the beta-decay measurements occurs near *noon* in or near *June*. Neutrinos detected near noon are traveling *towards* the Sun, not from the Sun. These can only be cosmic neutrinos.

This signal matches the expectation for the production of a cosmic influence such as dark matter as estimated by Freese (2017), and matches results of the DAMA/Libra dark matter experiment (Bernabei et al. 2018), which exhibits a strong peak in early June.

Figure 13 does not show an obvious enhancement at midnight, implying that the solar neutrino flux is smaller than the average neutrino flux by at least a factor of 10. Estimates of the solar neutrino production presented by Bahcall (1989) lead to an estimate for the solar neutrino flux at Earth of

$$F(\text{solar neutrinos}) = 10^{10.9} \ cm^{-2} \ s^{-1} \quad (8.1)$$

(mainly pp neutrinos). This suggests that we consider a cosmic neutrino flux of at least

$$F(\text{cosmic neutrinos}) = 10^{12} \ cm^{-2} \ s^{-1} \quad . \quad (8.2)$$

If neutrinos approach the Sun with negligible speed, the speed they will have at the Earth's orbit is given by

$$v(\text{infall}) = \left(\frac{2 G M_{solar}}{r}\right)^{1/2}. \qquad (8.3)$$

With $G = 10^{-7.18}$, $M = 10^{33.30}$ and $r = 1$ AU $= 10^{13.18}$ (all in cgs units), we find that

$$v(\text{infall}) = 10^{6.6} \text{ cm s}^{-1}. \qquad (8.4)$$

Hence if the local speed of cosmic neutrinos is due to the Sun's gravitational influence, the cosmic neutrino number density at the Earth's orbit may be estimated to be

$$n_v = F/v = 10^{4.5} \text{ cm}^{-3}. \qquad (8.5)$$

The cosmic neutrino mass density may then be estimated to be

$$\rho_v = 10^{4.5} m_v \text{ g cm}^{-3}, \qquad (8.6)$$

where $m_v$ is the neutrino mass in grams.

Gaitskell (2004) has estimated the density of cosmic dark matter to be

$$\rho_c = 10^{-28.73} h^2 \text{ g cm}^{-3}, \qquad (8.7)$$

where $h$ is the dimensionless form of the Hubble constant in units of 100 km/s/Mpc.

The current experimental estimate of $h$ is approx. 0.7, leading to an estimate of $10^{-29.0}$ $g\ cm^{-3}$ for the density of dark matter. We see from Equation (8.6) that this requires a mass of $10^{-33.5}$ g, i.e. 0.2 eV, for the neutrino mass.

This estimate is consistent with the upper limit of 1 eV for the neutrino mass, as determined by the Katrin experiment (Aker et al. 2019). It therefore appears that galactic neutrinos may be able to supply the mass of dark matter at least on a local scale.

## 9. Discussion

This article has been concerned with variability of the solar neutrino flux and the variability of certain nuclear decay processes, and evidence for a relationship between the two. The analysis presented in Section 6 gives overwhelmingly strong evidence for the variability of radon-decay measurements made by the Geological Survey of Israel, as is clear also from Figure 13, that shows the beta-decay measurement as a function of both date and hour of day.

The remarkably close agreement of an oscillation at 9.43 y$^{-1}$ of the solar neutrino flux (discussed in Section 4) and the same oscillation in GSI radon-decay measurements (discussed in Section 6) is strong evidence of a linkage between the two processes. This linkage becomes more obvious in two recent articles (Sturrock et al. 2021a, Sturrock et al, 2021b) which show that the agreement concerning the oscillation at 9.43 year$^{-1}$, evident in the power spectra formed from Super-Kamiokande measurements and from Geological Survey of Israel Measurements, is an agreement not only in power but also in amplitude and phase.

We also note that a triplet of oscillations separated by 1 year$^{-1}$ (approximately 7.43 year$^{-1}$, 8.43 year$^{-1}$ and 9.43 year$^{-1}$) and two doublets (11.43 year$^{-1}$ and 12.43 year$^{-1}$, and 12.65 year$^{-1}$ and 13.65 year$^{-1}$) are further strong evidence of a solar influence on the beta-decay process. (See Table 4.)

We note that 13.65 year$^{-1}$ is the sidereal rotation rate of the solar radiative zone (Schou et al., 1998), and that 12.65 year$^{-1}$ is the corresponding synodic rotation rate of that zone. The occurrence of both oscillations may be attributed to an internal rotation axis that differs from the normal to the ecliptic (Sturrock and Bai, 1992). The probability that the oscillation at 12.63 year$^{-1}$ could have occurred by chance in the solar-rotation band is estimated to be of order 10$^{-27}$.

However, evidence reviewed in Section 7 indicates that the relevant physical processes have *directional* characteristics. As we remarked in that section, *any variability may be attributable to variation in the direction of propagation of decay products, and whatever influences the directionality of decay products is related to the directionality of the ambient neutrino flux.*

This directionality relationship would be expected if neutrinos could directly influence the beta-decay process. However, such a direct influence seems unlikely in view of the exceedingly small cross-section for the influence of neutrinos on protons and electrons. (The corresponding cross-section is of order $10^{-46}$ cm$^2$; Bahcall 1989) *This suggests that some other particle or field* (which we refer to hypothetically as a "neutrello") *may be responsible for a coupling between neutrinos and beta decays.*

If the directionality of the decay products is in fact determined by the directionality of the ambient neutrino flux, this would seem to indicate that the coupling is a *collective* process rather than a particle-particle process. To appreciate the difference between a particle-particle interaction and a collective interaction, we may recall the difference in an electron-ion plasma (Sturrock, 1994, 9 - 14). The particle-particle interaction is simply the electrostatic force of one particle (electron or ion) on another particle. The collective interaction is attributable to the total electromagnetic field of all charged particles. In a plasma, the latter is far more significant than the former. Similarly, the combined influence of all neutrellos in the solar system may exceed the influence of any single neutrello by a very large factor.

However, many experiments that might – according to this scenario - show evidence of variability of beta-decays fail to do so (Kossert & Nahle, 2014, Pomme, 2016). In this connection we may note that, of the experiments mentioned in this article, the GSI experiment (which gives the strongest evidence of variability) is unique in that *the beta- decay process and the propagation of decay products to the detectors occur entirely in air.* The significance of this fact is borne out by the experiments of Bellotti (2015) that were discussed in Section 7. When the decay of radon and the propagation of decay products to the gamma detector occurred in air, the detector manifested diurnal oscillations, but when these processes occurred in olive oil, there was no evidence of diurnal oscillations.

These facts, taken together, suggest that variability of decay products may be attributed to *directionality*, such that the direction of travel of decay products is conserved if the relevant events occur in air, but the direction of travel of decay products is lost if the relevant events occur in olive oil. These considerations suggest that *the key criterion for the detectability of decay variations is the directionality of the decay process. If directionality is conserved, oscillations may be detected, but if directionality is lost, oscillations will not be detected.*

Our comparison, in Section 7, of results of the GSI experiment and of two experiments by Bellotti et al. (2015) gave further evidence of the close connection between the two physical processes of neutrino variability and beta-decay variability. An experiment by Steinitz et al. (2015) yields direct evidence for beta-decay directionality. This concept also explains the difference between GSI measurements made at noon and at midnight.

In retrospect, we can see that the interpretation of beta-decay variability as a manifestation of decay-rate variability was a red herring that jeopardized our understanding of neutrino variability and beta-decay variability and their relationship: counter-intuitively, variability of beta-decay products does not require variability of beta-decay rates.

The picture that emerges from these investigations is as follows: *Neutrinos do not influence whether or not decay occurs. However, the direction of emission of decay products is influenced by the ambient neutrino flux, if and when decay occurs.*

Some of the concepts we have introduced suggest experiments that may confirm or discredit these concepts. Some such experiments are suggested in Appendix C.

Concerning the concept of collectivity - that particles interact in a collective manner rather than through binary interactions - this concept can be checked by searching for evidence of a *correlation* between two experiments that are not co-located. One can start with small separations of only 2 or 3 feet. However, if one finds evidence for a correlation over a small distance, it would then be appropriate to increase the distance, up to considering experiments on different continents.

This article has been concerned with variability of the solar neutrino flux and the variability of certain nuclear decay processes, and evidence for a relationship between the two.

The analysis presented in Section 6 gives overwhelmingly strong evidence for the variability of radon-decay measurements made by the Geological Survey of Israel. The remarkably close agreement of an oscillation at 9.43 year$^{-1}$ of the solar neutrino flux (discussed in Section 4) and the same oscillation in GSI radon-decay measurements (discussed in Section 6) is strong evidence of a linkage between the two processes. This linkage become more obvious in two recent articles (Sturrock et al. 2021a, Sturrock et al, 2021b) which show that the agreement concerning the oscillation at 9.43 year$^{-1}$, evident in the power spectra formed from Super-Kamiokande measurements and from Geological Survey of Israel Measurements, is an agreement not only in power but also in amplitude and phase. Furthermore, a triplet of oscillations separated by 1 year$^{-1}$ (approximately 7.43 year$^{-1}$, 8.43 year$^{-1}$ and 9.43 year$^{-1}$) and two doublets (11.43 year$^{-1}$ and 12.43 year$^{-1}$, and 12.65 year$^{-1}$ and 13.65 year$^{-1}$) are further strong evidence of a solar influence on the beta-decay process.

Concerning the significance of the *transparency* of the medium:- One could explore the role of the medium - beginning with a repeat of the Bellotti (2015) experiment - to determine whether gamma measurements are a function of the *optical depth* of the source with respect to the detector.

Concerning the concept of *directionality*:- The first step is to see if the measurement is a function of the vector between the source and the detector (as in the GSI experiment). A simple test will be to set up two identical experiments – each comprising a single GM detector with a small amount of radioactive material on the entry window. Set up two such experiments, *collocated but pointing in different directions*. If these experiments yield significantly different results, one might set up an array of detectors around a single source to determine the response to directionality.

According to the results of this article, experiments that appear to involve collective processes comprise neutrino detectors that are far more sensitive than detectors that depend on binary processes.

*Acknowledgements*

The author is indebted to the Brookhaven National Laboratory, the GALLEX, GNO, Homestake and Superkamiokande Consortia, the Physikalisch-Technische Bundesanstalt, and the Geological Survey of Israel for access to data referred to in this article. He is indebted to Ephraim Fischbach for introducing him to the enigma of nuclear-decay variability, and to Oksana Piatibratova, Jeffrey Scargle, Felix Scholkmann, Gunther Walther, Mark Weber and Mike Wheatland for their generous collaboration. Many colleagues and correspondents have contributed valuable thoughts and advice, including Evgeni Akhmedov, Yuri Baurov, Roger Blandford, Timothy Groves, Dan Javorsek, Jere Jenkins, Karsten Kossert, Luciano Pandola, Fabian Pease, Vahe Petrosian, Joao Pulido, Gioacchino Ranucci, Roger Romani, Heinrich Schrader, Robert Wagoner and Teimuraz Zaqarashvili. Special thanks are due to Todd Hoeksema for preparing this article for submission, and to Alexander Parkhomov and an anonymous reviewer for their insightful reviews. This article is dedicated to the memory of the late Gideon Steinitz of the Geological Survey of Israel, who has been the outstanding leader in the study of nuclear-decay variability.

**Appendix A. A likelihood procedure for power spectrum analysis of unevenly spaced data.**

Beginning with Section 3, power-spectrum analysis has played a key role in our investigations. The current basic procedure for power spectrum analysis of data that may have irregular spacing is the Lomb-Scargle procedure (Lomb 1976; Scargle, 1982).

For some purposes, as for instance in Section 6, it is necessary to estimate the amplitude and phase of each oscillation, in addition to the power. Hoecke (1998) and Zeichmeister and Kurster (2009) have developed procedures for these calculations, extending the original calculations of Lomb and Scargle.

We have found it convenient to carry out similar calculations using complex-variable notation, so as to incorporate amplitude and phase. This leads to more compact formulas and their derivations, which are well suited for calculations in Matlab notation, generating estimates of power, amplitude and phase for each of a range of frequencies. This procedure also has the advantage that it can be applied, with only slight modification, to data (such as neutrino measurements) that are compiled in *bins* of specified finite duration. If necessary, it can also take account of the decay of measurements during each bin as in Sturrock et al. (1997).

We begin with the flux values $g_r$ and the error terms $\sigma_r$. It is convenient but not essential to normalize data, as below, so that amplitude measurements will denote depth of modulation:

$$x_r = g_r / mean(g_r) - 1 . \qquad (A.1)$$

With the notation

$$w_r = 1/\sigma_r^2 , \qquad (A.2)$$

we need to find K that minimizes the negative-log-likelihood (the "unlikelihood")

$$V_0 = \frac{1}{2} \sum_r w_r (x_r - K)^2 . \qquad (A.3)$$

When this function is at its minimum, it is stationary with respect to any small change in K, so that

$$\frac{\partial V_0}{\partial K} \equiv -\sum w_r (x_r - K) = 0. \tag{A.4}$$

Hence

$$K = \frac{\sum w_r x_r}{\sum w_r}. \tag{A.5}$$

We now write

$$x_r = K + Y_r. \tag{A.6}$$

We wish to fit this to the form

$$Y_r = A e^{i\omega t_r} + A^* e^{-i\omega t_r}. \tag{A.7}$$

We therefore form

$$V = \frac{1}{2} \sum w_r \left( y_r - A e^{i\omega t_r} - A^* e^{-i\omega t_r} \right)^2. \tag{A.8}$$

We note that V has its minimum value when it is stationary with respect to any small change in the complex variable A. We therefore set

$$\frac{\partial V}{\partial A^*} \equiv -\sum e^{-i\omega t_r} w_r \left( y_r - A e^{i\omega t_r} - A^* e^{-i\omega t_r} \right) = 0. \tag{A.9}$$

(So also for the complex conjugate.)
If we write

$$\begin{aligned} M_{11} &= \sum w_r, \quad M_{12} = \sum w_r e^{-2i\omega t_r}, \\ M_{21} &= M_{12}{}^*, \quad M_{22} = M_{11} \end{aligned} \tag{A.10}$$

$$\begin{aligned} Y_1 &= \sum w_r y_r e^{-i\omega t_r} \\ Y_2 &= Y_1{}^* \end{aligned} \tag{A.11}$$

$$M = \begin{pmatrix} M_{11} & M_{12} \\ M_{21} & M_{22} \end{pmatrix}, \tag{A.12}$$

$$Y = \begin{pmatrix} Y_1 \\ Y_2 \end{pmatrix}, \tag{A.13}$$

and

$$X = \begin{pmatrix} X_1 \\ X_2 \end{pmatrix} = \begin{pmatrix} A \\ A^* \end{pmatrix}, \tag{A.14}$$

then Equation (A.9) and its conjugate may be combined as

$$MX = Y \tag{A.15}$$

which inverts to give

$$X = M^{-1} \times Y. \tag{A.16}$$

With the value of A found in this way, we can evaluate V from (A.8). Hence we can evaluate the power from the increase in likelihood, given by

$$S = V_0 - V. \tag{A.17}$$

This calculation is available as a Matlab file.

**Appendix B. Research at Standards Laboratories**

The possibility that some nuclear processes may be variable is of some interest to analysts at standards laboratories, since one of their responsibilities is to determine the decay rates of radioactive nuclei.

Karsten Kossert and Ole Nahle (2015) at the Physikalisch Technische Bundesanstalt (PTB) in Braunschweig, Germany, have set up an experiment specifically to search for evidence of variability. The abstract of their article, entitled *Disproof of solar influence on the decay rates of 90Sr/90Y*, reads:

> *A custom-built liquid scintillation counter was used for long-term measurements of 90Sr/90Y sources. The detector system is equipped with an automated sample changer and three photomultiplier tubes, which makes the application of a triple-to-double coincidence ratio (TDCR) method possible. After decay correction, the measured decay rates were found to be stable and no annual oscillation could be observed. Thus, the findings of this work are in strong contradiction to those of Parkhomov who reported on annual oscillations when measuring the 90Sr/90Y with a Geiger-Muller counter. Sturrock et al (2012) carried out a more detailed analysis of the experimental data from Parkhomov, and claimed to have found correlations between the decay rates and processes inside the Sun. These findings are questionable, since they are based on inappropriate experimental data as is demonstrated in this work. A frequency analysis of our activity data does not show any significant periodicity.*

A possibly significant point is that this experiment differs from most standards-laboratory experiments (but is similar to the BNL experiment) in that it involves an automated sample changer, which necessarily leads to an air gap between the specimen and the detector. (Most experiments at standards laboratories do not contain an air gap.)

Kossert and Nahle generously made their data available to Sturrock and his colleagues for independent analysis. This review (Sturrock, et al. 2016) uncovered a glitch in the Kossert-Nahle calculations. When corrected, the Kossert-Nahle data were found to exhibit a significant oscillation close to 11 year$^{-1}$. An analysis of GSI measurements for the exact time interval selected by Kossert and Nahle revealed a strong oscillation at 11.35 year$^{-1}$, as found in the power-spectrum analysis of Section 6.

Stefan Pomme of the European Joint Research Center in Geel, Belgium, who has taken a keen interest in this topic, has published many relevant articles. We here comment briefly on one of his most recent articles (Pomme, 2019) which lists references to 10 of his earlier publications.

The title of this article is *Solar influence on radon decay rates: irradiance or neutrinos?* and the abstract reads

> *Radon decay rate data from 2007 to 2011, measured in a closed canister with one gamma counter and two alpha detectors, were made available for analysis by the Geological Survey of Israel (GSI). Sturrock et al. have published several papers in which they claim that decay rate variations in the gamma counter can be associated with solar rotation. They assert influences by solar and cosmic neutrinos on beta decay and draw unsubstantiated conclusions about solar dynamics. This paper offers an alternative explanation by relating the daily and annual patterns in the radon decay rates with environmental conditions. Evidence is provided that the radon measurements were susceptible to solar irradiance and rainfall, whereas there is no indication that radioactive decay is influenced by the solar neutrino flux. Speculations about solar dynamics based on the concept of neutrino-induced beta decay are ill-founded.*

The possible role of environmental influences is important, and Pomme' advances interesting speculations on the possible influence of local temperature, rainfall etc., on GSI gamma measurements. Fortunately, the GSI experiment provides measurements of the internal temperature and pressure (and line voltage) on the same schedule as the decay-product measurements. The relationship of gamma measurements to temperature, pressure and voltage has been examined, in the form of spectrograms, in a recent article (Sturrock et al. 2018). The results are displayed in Figure 16 of that article.

In brief, there is no correspondence between decay data and environmental data. The amplitude of rotation-band oscillations in gamma measurements is much greater than that of temperature, pressure or voltage. The gamma plot shows highly significant and highly localized (all nighttime) features, the strongest at midnight at 11.4 year$^{-1}$. The temperature plot shows only significantly *weaker* features, in *daytime* data, at a number of frequencies *(none at 11.4 year$^{-1}$)*.

Since environmental conditions inside the experiment have no discernible influence on beta-decay measurements inside the experiment, it would seem highly unlikely that environmental measurements outside the experiments would have any influence on nuclear processes inside the experiment. Hence study of Kossert and Nahle's experiment and of Pomme's analysis of GSI data provides supporting evidence for the variability of nuclear decays.

**Appendix C. Proposed experiments**

There are relevant theoretical ideas that it would be good to check experimentally.

One is the concept of *collectivity* - that particles interact in a collective manner rather than through binary interactions.

This can be checked by searching for evidence of a *correlation* between two experiments that are not co-located. One can start with small separations of only 2 or 3 feet. However, if one finds evidence for a correlation over a small distance, it would then be appropriate to increase the distance, up to considering experiments on different continents.

The *transparency* of the medium.

One could also explore the role of the medium, beginning with a repeat of the Bellotti (2015) experiment, to determine whether gamma measurements are a function of the optical depth (the line integral of the absorption coefficient) of the source with respect to the detector.

Another concept to check is that of *directionality*.

The first step is to see if the measurement is a function of the vector between the source and the detector (as it is in the GSI experiment). To start with, one can use the simplest test. Set up two identical experiments – each comprising a single GM detector with a small amount of radioactive material on the entry window. Set up two such experiments, collocated but pointing in different directions. Then repeat, one pair (source and detector) with a short distance between source and detector. If these experiments yield significantly different results, set up an array of detectors around a single source, so as to derive a *polar diagram*.

*Acknowledgements*

The author is indebted to the Brookhaven National Laboratory, the GALLEX, GNO, Homestake and Superkamiokande Consortia, the Physikalisch-Technische Bundesanstalt, and the Geological Survey of Israel for access to data referred to in this article. He is indebted to Ephraim Fischbach for introducing him to the enigma of nuclear-decay variability, and to Oksana Piatibratova, Jeffrey Scargle, Felix Scholkmann, Gunther Walther, Mark Weber and Mike Wheatland for their generous collaboration. Many colleagues and correspondents have contributed valuable thoughts and advice, including Evgeni Akhmedov, Yuri Baurov, Roger Blandford, Timothy Groves, Dan Javorsek, Jere Jenkins, Karsten Kossert, Luciano Pandola, Fabian Pease, Vahe Petrosian, Joao Pulido, Gioacchino Ranucci, Roger Romani, Heinrich Schrader, Robert Wagoner and Teimuraz Zaqarashvili. Special thanks are due to Todd Hoeksema for preparing this article for submission, and to Alexander Parkhomov and an anonymous reviewer for their insightful reviews. This article is dedicated to the memory of the late Gideon Steinitz of the Geological Survey of Israel, who has been the outstanding leader in the study of nuclear-decay variability.

Table 1. Power estimates of oscillation at 9.43 year$^{-1}$ from several analyses of the Super-Kamiokande data (Sturrock, Caldwell, et al. (2005); Sturrock & Scargle, 2005).

| Analysis procedure | Power |
|---|---|
| Lomb-Scargle, using mean times of bins | 5.90 |
| Lomb-Scargle, using mean live times of bins | 6.18 |
| Likelihood, using start and end times | 11.51 |
| Likelihood, using start and end times, and mean live times | 11.67 |
| Likelihood, using start and end times, mean live times, and asymmetrical error terms | 13.24 |

Table 2. The top 10 peaks in the frequency band 6 – 16 year$^{-1}$ in the power spectrum derived from Super-Kamiokande measurements by a likelihood procedure that takes account of the start time and end time of each bin, the flux estimate at the mean live time, and the upper and lower error estimates.

| Frequency (year$^{-1}$) | Power | Order |
|---|---|---|
| 8.29 | 5.60 | 3 |
| 8.74 | 2.97 | 10 |
| 8.98 | 3.91 | 6 |
| 9.43 | 13.24 | 1 |
| 10.68 | 3.51 | 7 |
| 11.29 | 3.34 | 8 |
| 12.31 | 6.24 | 2 |
| 12.69 | 4.79 | 4 |
| 14.87 | 3.09 | 9 |
| 15.72 | 4.79 | 5 |

Table 3. Top 20 peaks in the power spectrum formed from GSI noon data in the frequency band 6 – 16 year$^{-1}$.

| Frequency (year$^{-1}$) | Power | Order |
|---|---|---|
| 6.07 | 4.4 | 16 |
| 6.72 | 4. 5 | 15 |
| **7.45** | **10.7** | **2** |
| 7.81 | 7.8 | 5 |
| 7.96 | 3.5 | 20 |
| **8.47** | **4.1** | **17** |
| 8.85 | 6.5 | 7 |
| 9.21 | 4.6 | 13 |
| **10.31** | **5.0** | **11** |
| 10.74 | 6.4 | 8 |
| 10.90 | 5.9 | 10 |
| **11.34** | **14.9** | **1** |
| 12.37 | 3.7 | 19 |
| **12.65** | **6.8** | **6** |
| 12.86 | 7.9 | 4 |
| 13.13 | 9.6 | 3 |
| **13.67** | **6.0** | **9** |
| 14.14 | 4.9 | 12 |
| 14.99 | 3.7 | 18 |
| 15.24 | 4.5 | 14 |

Table 4. Top 20 peaks in the power spectrum formed from GSI midnight data in the frequency band 6 – 16 year$^{-1}$.

| Frequency (year$^{-1}$) | Power | Order |
|---|---|---|
| 6.13 | 18.5 | 19 |
| 7.18 | 18.9 | 18 |
| **7.45** | **20.7** | **15** |
| 7.80 | 37.1 | 5 |
| 8.30 | 22.2 | 14 |
| **8.46** | **42.4** | **4** |
| 8.87 | 19.6 | 16 |
| 9.21 | 24.8 | 12 |
| **9.44** | **22.6** | **13** |
| 9.95 | 18.2 | 20 |
| 10.93 | 36.4 | 7 |
| **11.35** | **65.5** | **1** |
| 11.91 | 19.1 | 17 |
| **12.35** | **31.7** | **9** |
| **12.63** | **61.4** | **2** |
| 12.86 | 32.2 | 8 |
| **13.67** | **31.1** | **10** |
| 13.90 | 25.4 | 11 |
| 14.14 | 37.1 | 6 |
| 15.00 | 51.3 | 3 |

SSR2021B_Figs_211012.doc

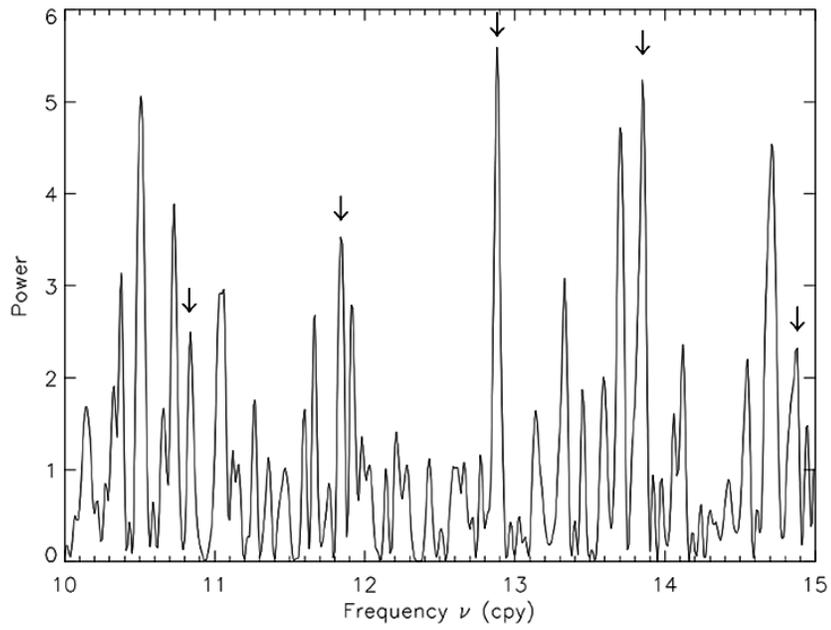

Figure 1. Power spectrum formed from Homestake flux measurements. [HomestakeDG200222.eps]

**48**

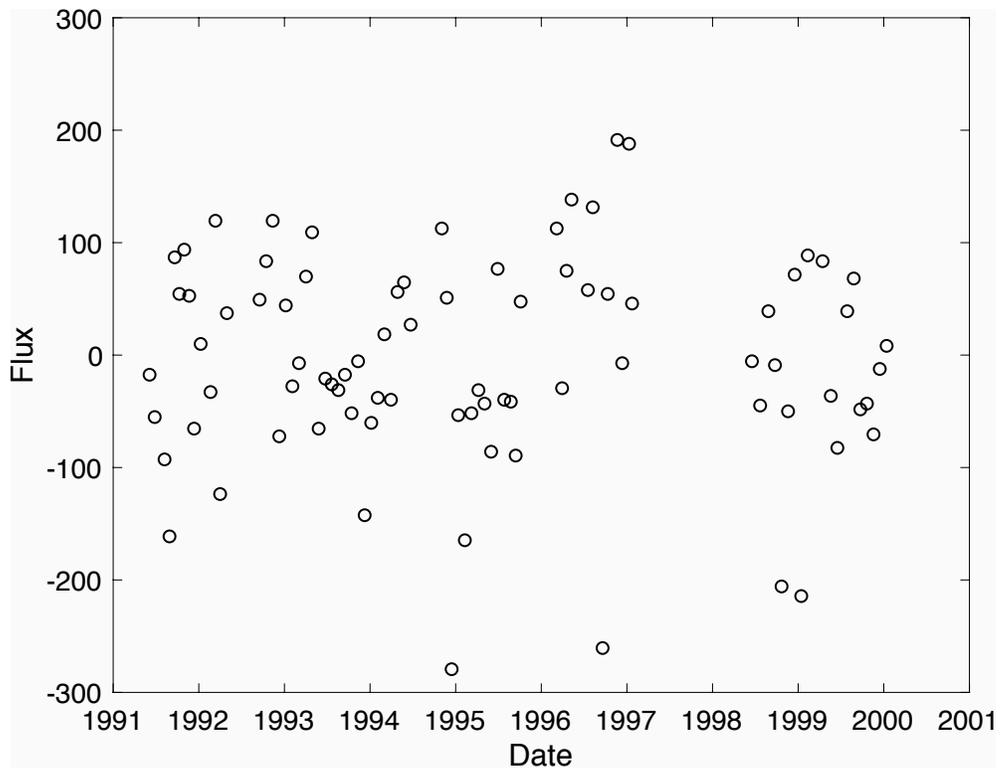

Figure 2. Flux versus date for GALLEX-GNO measurements.

[Neu01G02a, Neu21D01.eps]

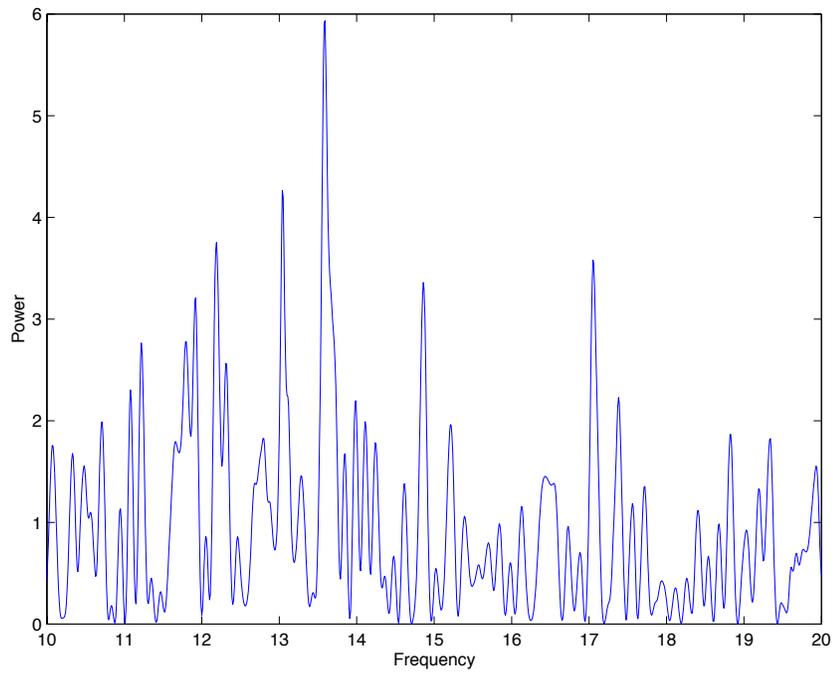

Figure 3. Power spectrum formed from GALLEX-GNO flux measurements. The peak power is 5.93 at 13.59 year$^{-1}$.   [Neu00K41.m??]

**50**

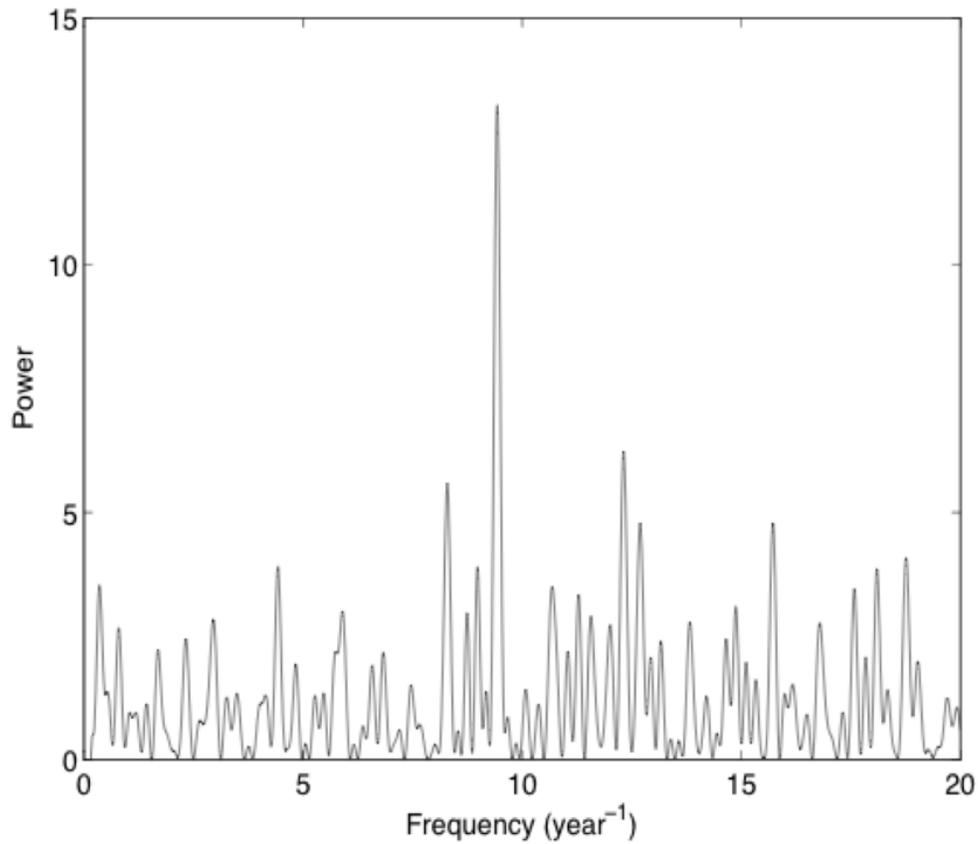

Figure 4. Power spectrum formed from Super-Kamiokande data, taking account of start time, end time, flux estimate, lower error estimate and upper error estimate (Sturrock and Scargle, 2008). [Decs18ZA02a0.pdf]

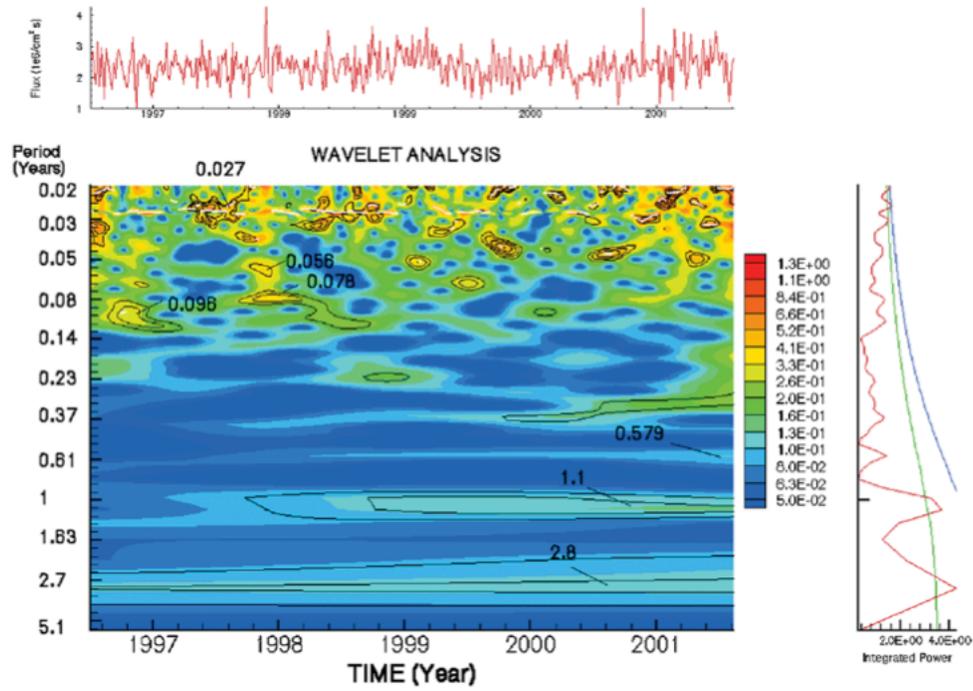

Figure 5. Local wavelet map of the Super-Kamiokande 5-day data [Figure 4 of Ranucci and Sello (2007)].

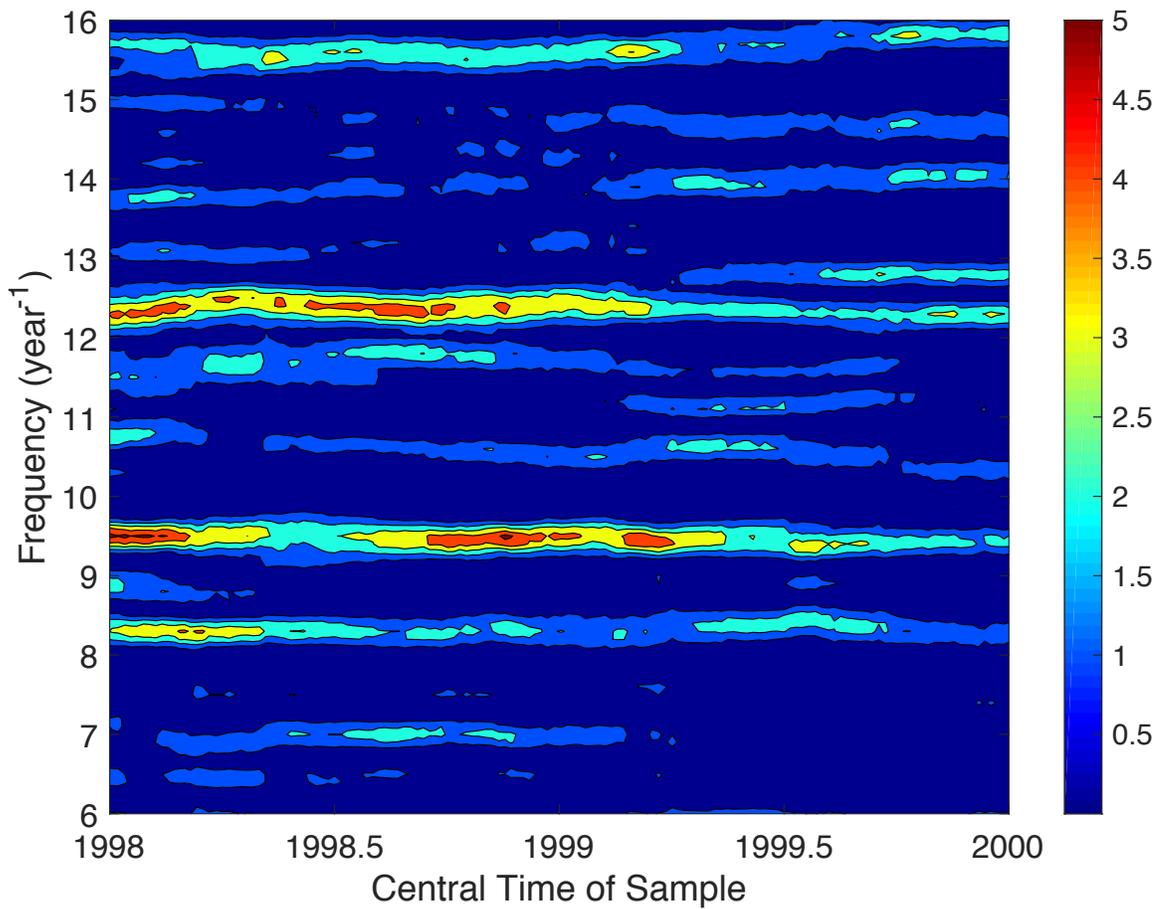

Figure 6. Spectrogram formed from Super-Kamiokande measurements. This spectrogram, derived from Super-Kamiokande measurements, is formed by power-spectrum analysis of a sequence of sections of data, each section of length 1,000 days. The principal oscillations are at 12.6 year$^{-1}$, 9.4 year$^{-1}$ and 8.4 year$^{-1}$. [Neu12G09k]

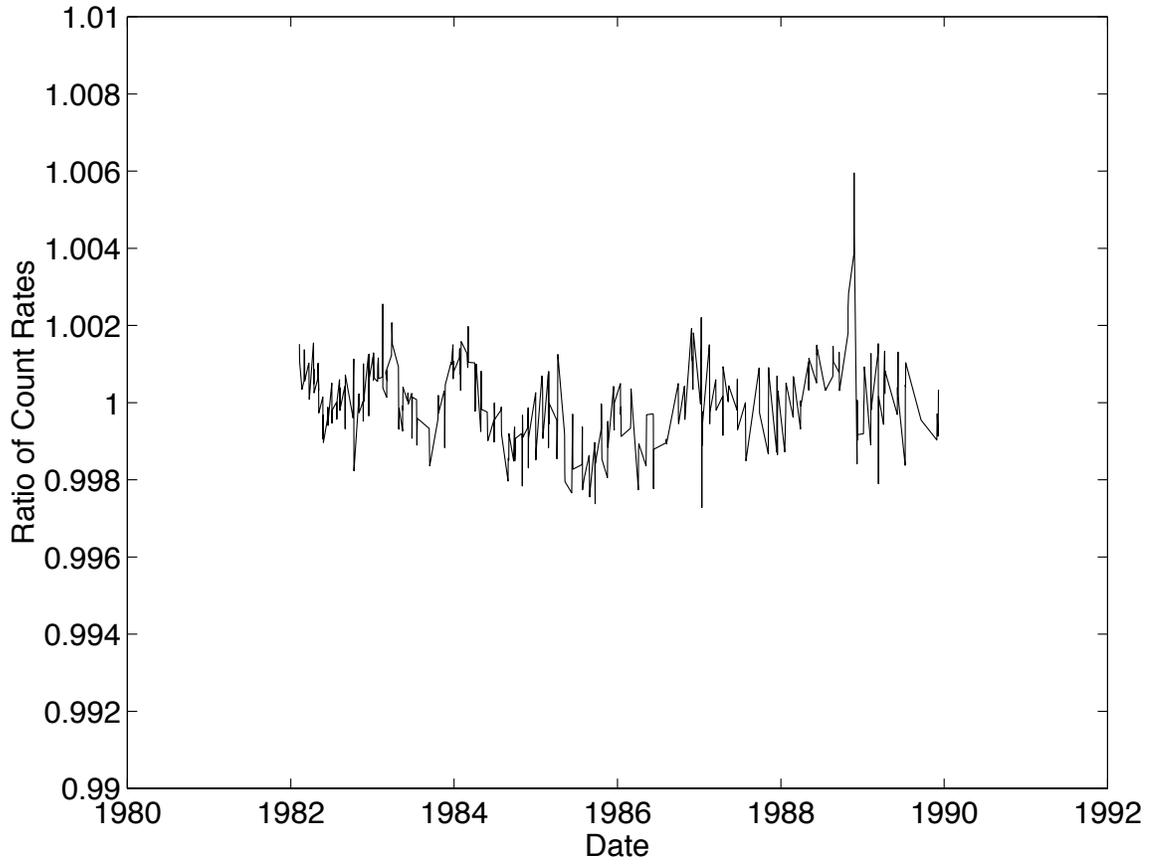

Figure 7. Ratio of the normalized count rate from the $^{32}$Si source to the normalized $^{36}$Cl source in athe BNL experiment. [Decs09ZZE01c]

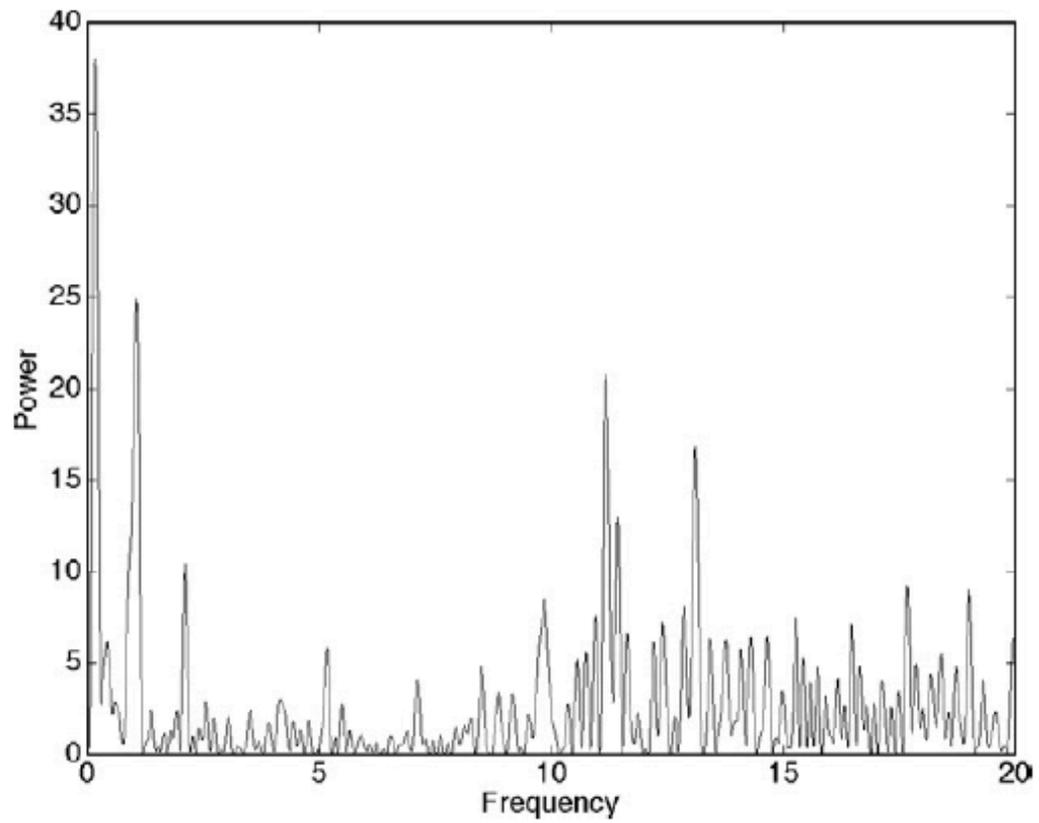

Figure 8. Power spectrum formed by a likelihood procedure from the ratio of the $^{32}$Si and $^{36}$Cl count rates. [Decs09ZZE01c0.eps, Neu00K41.m]

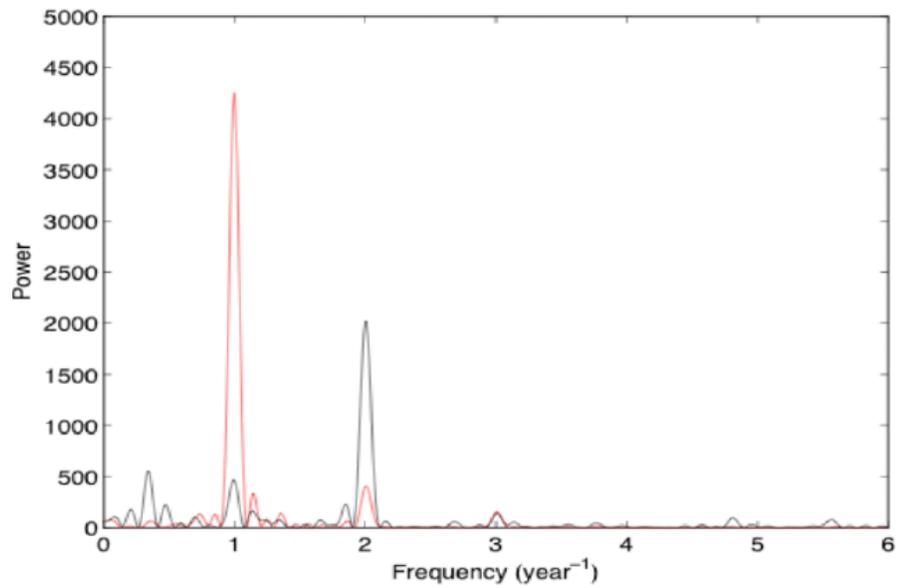

Figure 9. Power spectra formed from the 4-hour band of measurements centered on noon (red) and midnight (blue) for the frequency range 0 - 6 year$^{-1}$. We see that the strongest daytime oscillation is at 1 year$^{-1}$; the strongest nighttime oscillation is at 2 year$^{-1}$.

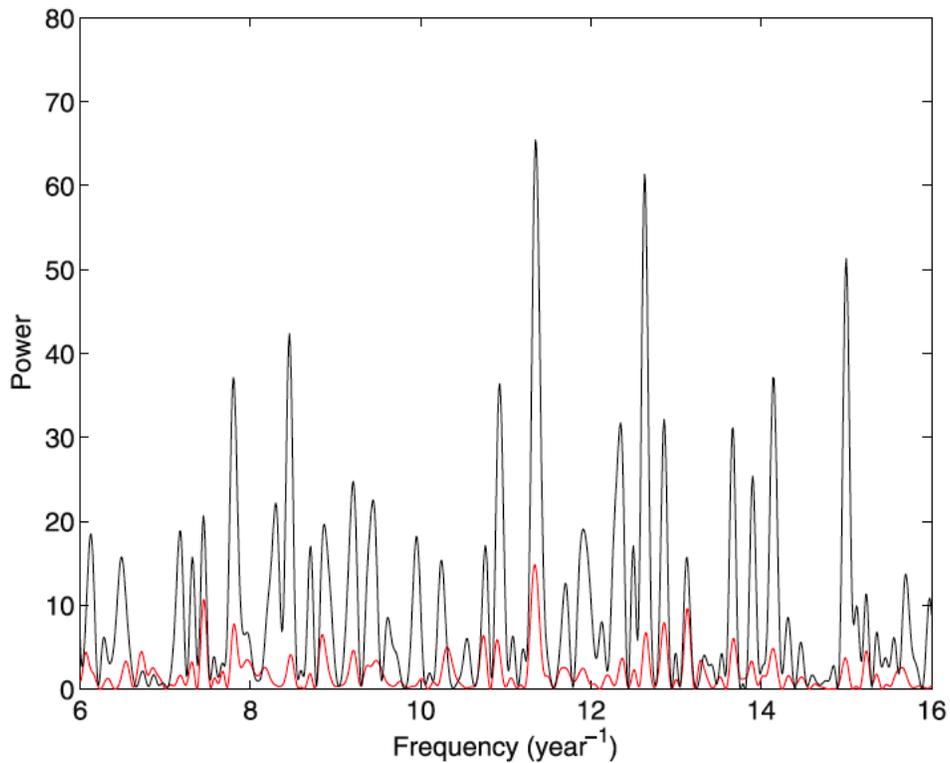

Figure 10. Power spectra formed from the 4-hour band of measurements centered on noon (red) and on midnight (blue) for the frequency band 6 - 16 year$^{-1}$. We see that there are strong oscillations in the candidate rotational frequency band 10 – 14 year$^{-1}$ (note especially the peaks at 11.35 year$^{-1}$ and at 12.65 year$^{-1}$) in the midnight data, but comparatively small oscillations in the daytime data. (Cf. Tables 6-1 and 6.2.)

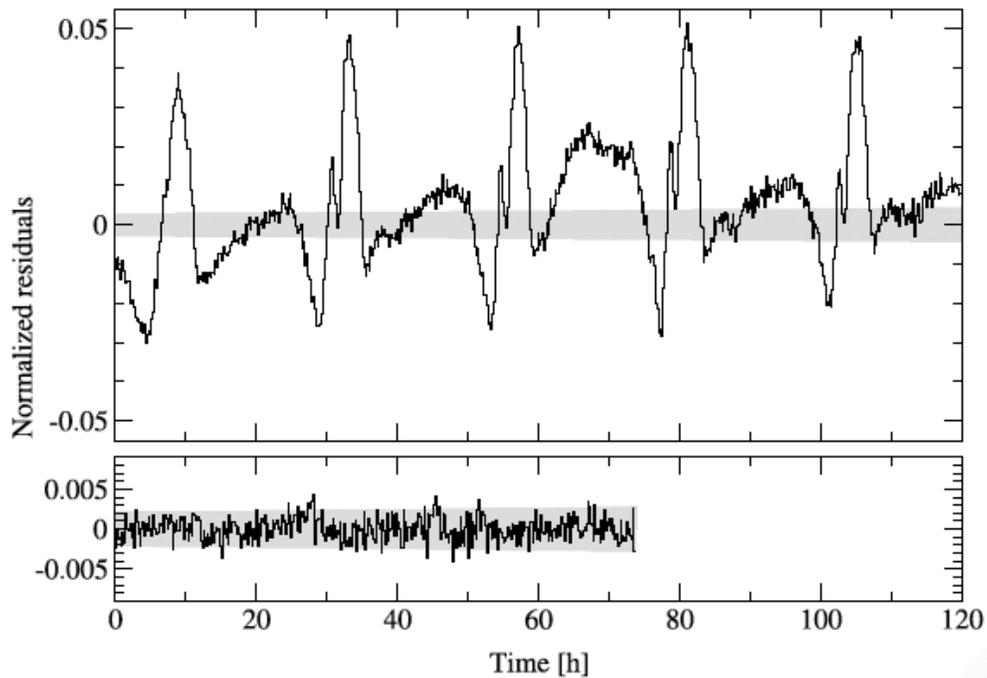

Figure 11. Normalized deviations from the exponential trend of the measured count rate as a function of time for radon in air: without (upper panel) and with (lower panel) the polystyrene particles inside the glass sphere containing air charged with radon. The shaded areas limit the region where fluctuations are within 2σ[1].

[1](Belotti 2015).

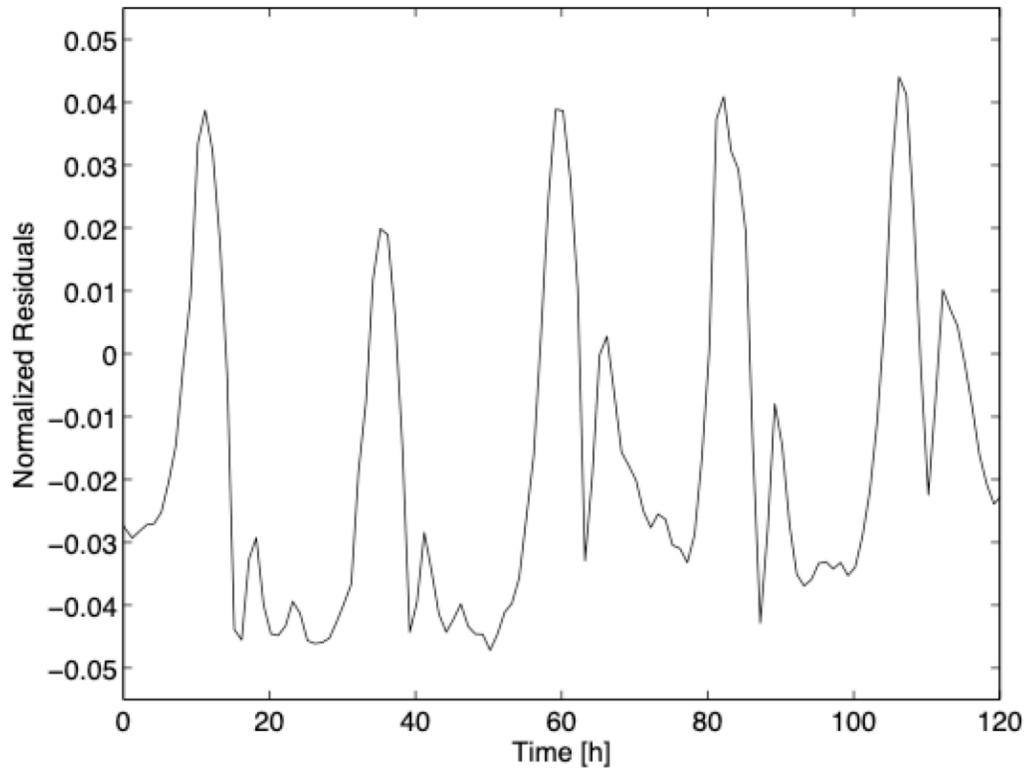

Figure 12. Normalized deviations from the exponential trend of the measured count rate as a function of time for the GSI experiment.

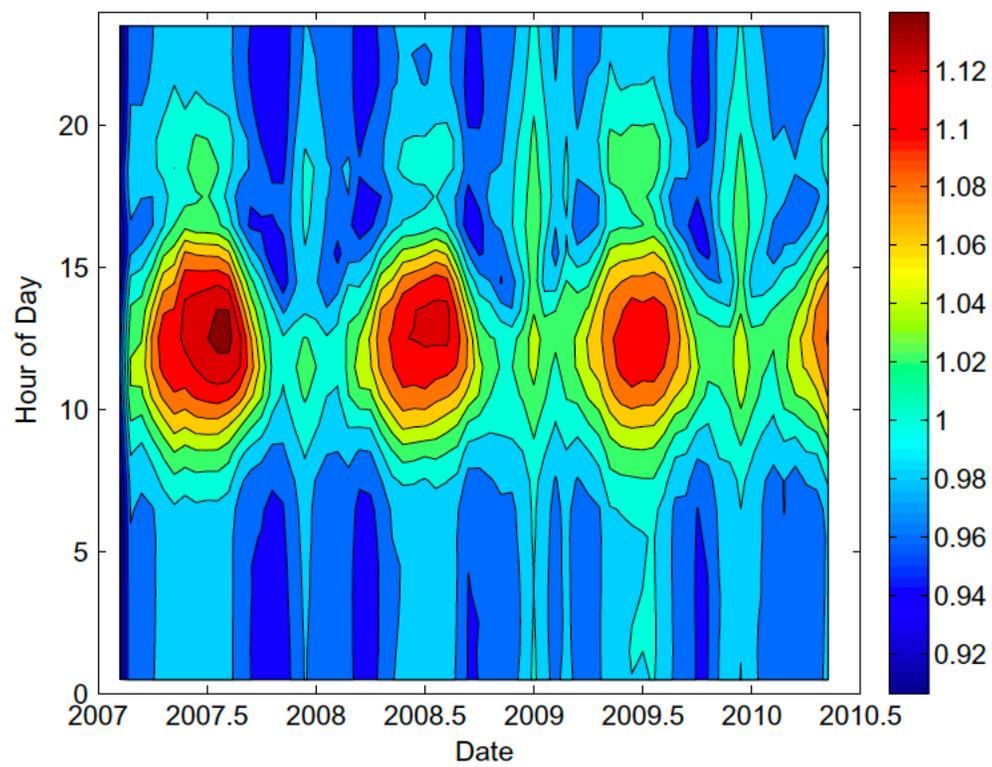

Figure 13. Normalized GSI gamma measurements as a function of date and hour of day.